\documentclass[11pt,letterpaper]{article}
\usepackage[margin=1in,headheight=14pt]{geometry}
\usepackage[utf8]{inputenc}
\DeclareUnicodeCharacter{2212}{\ensuremath{-}}
\usepackage[T1]{fontenc}
\usepackage{mathpazo}
\usepackage{amsmath,amssymb}
\usepackage{microtype}
\usepackage{longtable,booktabs,array,calc}
\usepackage{caption,threeparttable,placeins}
\newcommand{\Needspace}[1]{\pagebreak[2]}

\usepackage{graphicx}
\usepackage[round,authoryear]{natbib}
\usepackage{xurl}
\usepackage{xcolor}
\definecolor{midblue}{RGB}{0,70,140}
\usepackage[colorlinks=true,allcolors=midblue]{hyperref}
\hypersetup{
  pdftitle={Is the business of frontier AI companies selling tokens? Value capture and equity prices},
  pdfauthor={Andrew J. Peterson},
  pdfsubject={AI governance},
  pdfkeywords={Artificial intelligence; firm valuation; business models; value capture; firm boundaries; technology licensing}
}
\usepackage{fancyhdr}
\makeatletter
\newcommand{\sectionprefix}{}

\renewcommand{\@seccntformat}[1]{\csname #1prefix\endcsname\csname the#1\endcsname.\quad}
\newcommand{\manuscriptappendix}{%
  \appendix
  \renewcommand{\sectionprefix}{Appendix~}%
  \renewcommand{\thesubsection}{\thesection\arabic{subsection}}%
  \setcounter{table}{0}%
  \renewcommand{\thetable}{\thesection\arabic{table}}%
  \setcounter{figure}{0}%
  \renewcommand{\thefigure}{\thesection\arabic{figure}}%
  \renewcommand{\theHtable}{\thesection.\arabic{table}}%
  \renewcommand{\theHfigure}{\thesection.\arabic{figure}}%
}
\makeatother
\let\standardlabel\label
\title{Is the business of frontier AI companies selling tokens? Value capture and equity prices}
\author{Andrew J. Peterson\thanks{Assistant Professor, University of Poitiers. \href{mailto:andrew.peterson@univ-poitiers.fr}{andrew.peterson@univ-poitiers.fr}. ORCID: \href{https://orcid.org/0000-0002-0811-3515}{0000-0002-0811-3515}.}}
\date{October 8, 2026}

\begin{document}
\maketitle
\begin{abstract}
We examine business models that could support frontier AI companies’ valuations and the public evidence of plans to earn income beyond selling model access. 
Treating the announced financing prices as hypothetical operating-value targets, reverse cash-flow calculations for OpenAI and Anthropic yield combined year-ten revenue requirements of \$2.79 trillion or \$733 billion, depending on financial assumptions. 
Public documents identify proposed payments and ownership interests that could supply additional income, such as through partnerships on pharmaceuticals, robotics, or materials science.  %
We discuss how these arrangements relate to firm boundaries and value capture.
\end{abstract}

\vspace{10pt}
\begin{center}
firm boundaries; technology licensing.
\end{center}
\vspace{10pt}

\section{Introduction}\label{sec:introduction}

The valuations of frontier AI companies raise a question: what will these firms sell, and how will they earn the cash flows needed to support those prices? 
We take the announced valuations as targets for a conditional analysis, leaving open whether investors have correctly assessed the firms' prospects. 
Our research therefore asks whether model access sales support the announced prices under plausible cash-flow conditions, and if not, what other scenarios could better complete the revenue model. 
Frontier model providers can also directly perform work for customers or directly own or partner on businesses that develop and sell products, such as pharmaceutical, materials science, or robotic innovations.

To give an initial sense of the scale, OpenAI announced in March 2026 funding that put its post-money equity price at \$852 billion, and reported approximately \$2 billion in monthly revenue. 
Anthropic announced funding in May 2026 at a valuation of \$965 billion and its annualized revenue run rate above \$47 billion \citep{openai2026funding,anthropic2026h}. 
These are some of the world's most rapidly growing and dynamic companies in history, so there is a fair amount of uncertainty about their future path. 
Higher financing and IPO valuations being discussed also motivate testing how the cash-flow requirements change when only the valuation target rises \citep{bloomberg2026uaefunds,ft2026anthropicipo}. 
We subject these valuations to traditional financial analysis and find that supporting them requires strong assumptions about sustained revenue growth and future cash margins. 
We also consider whether additional sources of income could help explain those valuations.

We call subscriptions, business seats, and API-based model usage
the `access-and-seats baseline', or just `selling access'. 
This includes tools such as coding assistants when people pay through these channels, but excludes advertising, 
helping customers install and adapt AI systems, charging for a completed result, licensing intellectual property, or 
earning income from selling products, receiving royalties, or holding ownership stakes in businesses that commercialize AI-assisted discoveries.

First, we consider what must be necessary for access sales to suffice. 
The conditions are demanding but possible: a ten-year reverse valuation with a 15\% discount rate and a 25\% mature cash margin requires annual issuer revenue of approximately \$1.32 trillion for OpenAI and \$1.47 trillion for Anthropic. 
If we instead assume a 10\% discount rate and a 40\% margin, it would require \$347 billion and \$386 billion in annual revenues, respectively.

A second possibility we consider is thus that part of the investors' modeling is based on additional sources of income or financial value, which we only observe based on limited public signals. 
As one example, AI companies could collaborate with pharmaceutical firms on drug discovery and development, receiving research funding, payments for reaching agreed milestones, and royalties on resulting drug sales. Alternatively, they could share development costs and profits or participate through a jointly owned venture.  
OpenAI's CFO has proposed licensing and outcome payments, and OpenAI has announced an ownership stake in Thrive Holdings \citep{friar2026business,openai2025thrive}.
The stake could provide investment returns through distributions or realized gains on a sale. 
Scientific collaborations and information on internal research programs, which we document below, support the idea that these could play a part in the valuation scenarios, even if these ventures and markets may be speculative and the public announcements do not include details about the AI companies' specific financial arrangements.

The broader economic question is how the gains from AI are divided among frontier model providers, the firms and workers that use such models or are replaced by them, and consumers. 
Model providers may not capture the returns even when they develop a highly productive technology. 
Firms that control assets needed to commercialize the technology, such as manufacturing capacity or distribution networks, may capture some of the profits instead \citep{teece1986}.

We analyze public documents to see what arguments the funders give to justify their investments, along with any financial rights or commercial arrangements they announce. 
We also calculate the revenue and cash margins needed to support the announced valuations under explicit assumptions. 
This helps clarify whether the valuations rest on growth in access sales alone or also on expected income from other businesses, 
although of course we cannot establish the expectations investors actually held about these scenarios, recursive self-improvement (RSI), or other considerations. 
Section~\ref{sec:financing} establishes the financing inputs. 
Section~\ref{sec:framework} defines the framework and evidence design. 
Section~\ref{sec:financial-conditions} presents the financial conditions and strongest access-and-seats-based sales case. 
Section~\ref{sec:commercial-routes} examines additional payment and ownership routes,
and Section~\ref{sec:conclusion} concludes.

\section{Funding-round valuations and reported revenue}\label{sec:financing}

As a starting point, we ask how much revenue these companies would need to generate cash flows consistent with their funding-round valuations. %
We focus on OpenAI's March 2026 and Anthropic's May 2026 funding-round valuations and reported revenue figures. 
Table~\ref{tbl:financing} also shows earlier rounds for context.

\begin{table}[htbp]
\centering
\captionsetup{labelfont=bf,textfont=bf,justification=raggedright,
  singlelinecheck=false,skip=8pt}
\renewcommand{\arraystretch}{1.12}
\begin{threeparttable}
\caption{Financing announcements and reported sales}\label{tbl:financing}
\begin{tabular}{@{}
  >{\raggedright\arraybackslash}p{\dimexpr 0.2800\linewidth - 1.6800\tabcolsep\relax}
  >{\raggedleft\arraybackslash}p{\dimexpr 0.2400\linewidth - 1.4400\tabcolsep\relax}
  >{\raggedleft\arraybackslash}p{\dimexpr 0.2200\linewidth - 1.3200\tabcolsep\relax}
  >{\raggedright\arraybackslash}p{\dimexpr 0.2600\linewidth - 1.5600\tabcolsep\relax}@{}}
\toprule
\textbf{Announcement} & \textbf{Funding} &
\textbf{Post-money equity price} & \textbf{Reported sales pace} \\
\midrule
OpenAI, Mar 2025 & \$40B announced & \$300B & None in release \\
Anthropic Series F, Sep 2025 & \$13B completed & \$183B & August run
rate \textgreater\$5B/year \\
Anthropic Series G, Feb 2026 & \$30B raised & \$380B & \$14B/year run
rate \\
OpenAI, Mar 2026 & \$122B committed & \$852B & About \$2B/month \\
Anthropic Series H, May 2026 & \$65B raised & \$965B & Run rate
\textgreater\$47B/year \\
\bottomrule
\end{tabular}
\begin{tablenotes}[flushleft]
\footnotesize
\setlength{\itemsep}{3pt}
\item[] \textit{Notes.} Nominal US dollars. OpenAI's \$122B is committed
capital; Anthropic's \$65B includes \$15B of previously committed
hyperscaler investment. Anthropic's release dates the sales observation
only to ``earlier in May.''
\item[] \textit{Sources.} Company announcements
\citep{openai2025funding,anthropic2025f,anthropic2026g,openai2026funding,anthropic2026h}.
\end{tablenotes}
\end{threeparttable}
\end{table}

\FloatBarrier

Bloomberg has reported that OpenAI is seeking \$30 billion at a
pre-money valuation of about \$1.4 trillion, or about \$1.43 trillion post-money. 
At the time of writing, the round was still under negotiation
\citep{bloomberg2026openaitarget,bloomberg2026uaefunds}. 
Because this price is not final, we treat it as an alternative scenario, under which we recompute OpenAI's required revenue at the reported price, while the main
calibrations continue to use the valuations and sales from OpenAI's March
round and Anthropic's May round.\footnote{For Anthropic, we also test a \$2 trillion target based on reported investor IPO expectations, rather than an agreed offering price.
Anthropic's June~1 announcement of its confidential filing left the share count and offering
price open \citep{anthropic2026s1}. The \emph{Financial Times} attributed expectations of an Anthropic IPO valuation of \$2 trillion or more to several investors \citep{ft2026anthropicipo}, and Reuters reported on October~6 that a valuation above \$2 trillion was possible \citep{reuters2026anthropicipo}.  Reuters also reported a company revenue forecast of \$190-200 billion for 2028 and investors' use of forward revenue multiples \citep{reuters2026anthropicforecast}; we treat this forecast as context but do not directly use it as an input starting revenue.}

A funding announcement's headline value is not necessarily the value of the
business's operations. 
Preferred shares may carry protections absent from other classes, so applying their price to every share can overstate aggregate equity value \citep{gornall2020}. 
Some investors also supply computing services to the model providers, so they may gain from those sales as well as from their equity stakes \citep{openai2026amazon,anthropic2026h}.
Cash, liabilities, and future financing further separate equity value from the value of operations. 
SoftBank's February 27 agreement disclosed a \$30 billion OpenAI investment in preferred shares at a \$730 billion pre-money valuation, and adding the later \$122 billion round to that valuation yields the \$852 billion headline.
SoftBank's announcement, however, describes only its own investment, not the full capital structure \citep{softbank2026followon,softbank2026thirdtranche}. 
We therefore use the reported valuation only as a hypothetical
operating-value target and vary that target in a sensitivity analysis.

The reported revenue figures need careful interpretation. 
OpenAI's roughly \$2 billion in monthly revenue implies an annualized pace of \$24 billion.
Anthropic reported an annualized run rate above \$47 billion but gave no exact figure,
so we use that disclosed lower bound as the starting value. 
Relative to these run rates, the valuations imply ratios of 35.5 for OpenAI and
at most about 20.5 for Anthropic. 
These ratios offer rough context only; they are not directly comparable
price-to-sales multiples. 
Neither denominator is audited annual revenue, and reporting suggests that firms differ in whether cloud-mediated sales are counted gross or net \citep{axios2026accounting}. 
Public information is not enough to make a consistent adjustment, so we use the
figures as reported.

Furthermore, the announcement dates may differ from negotiation dates, and different information may be available to investors at different times. 
SoftBank's agreement preceded OpenAI's March release, while several commercial disclosures discussed below followed the relevant funding rounds.
Those later disclosures help identify possible business models,
but do not establish what investors knew when agreeing to the valuations.
Appendix~\ref{app:selection} records the announcements and their timing.

\section{Analytical approach and public evidence}\label{sec:framework}

Our goal is to try to deduce what model providers' valuations can tell us about potential business model paths. 
We combine a reverse cash-flow analysis with consideration of public documents that point to other potential sources of income or financial value. 
Beyond API and subscriptions, model providers may hope to capture more of the value chain. 
Whether they manage to do so or not depends on the structure of the market, such as who controls the complementary assets needed to commercialize a technology \citep{teece1986,teece2018}. 
Model providers could also license their technology while other firms
undertake downstream commercialization \citep{arora2001}.
A model provider whose technology helps develop a successful drug, for example, might earn only model-access (API) fees, or might also receive milestone payments, royalties, or returns on an ownership stake.

We focus on official OpenAI and Anthropic announcements from January 2025 through October 6, 2026 on financing, ventures, ownership interests, and payment or distribution rights, as well as investor statements, transaction reporting, and scientific collaborations to provide interpretation and context. 
Appendix~\ref{app:selection} records an inventory of these sources along with our selection rules and chronology.

For each possibility, we record the proposed business model, 
any disclosed payment or ownership rights, and reported commercial results,
including collaborations where the provider's financial role is unspecified. 
The documents do not provide the projections needed to value each business model separately, so our calculations use explicit financial assumptions. 
Appendix~\ref{app:methods} provides these claim classifications and sourcing information.

\section{Financial conditions and the access-sales case}\label{sec:financial-conditions}

Taking the announced valuations (Section~\ref{sec:financing}) as given, we solve for ten year revenue whose implied free cash flows, once discounted, equal each firm's announced valuation \citep{mauboussin2021,cogliati2011}. 
Here, free cash flow is revenue after operating costs, taxes, capital spending, and working-capital needs.

The assumed cash margins also include contractual payments to commercial partners. 
OpenAI's April 2026 Microsoft amendment preserves OpenAI-to-Microsoft
revenue-share payments through 2030, subject to a total cap, while ending
Microsoft-to-OpenAI revenue sharing \citep{openai2026microsoft}.

Let $P$ be the announced post-money equity price and  $V=\lambda P$ the hypothetical operating target. 
Let $R_0$ denote the annualized sales-pace proxy, $R_T$ the annual issuer revenue in year $T$, $m_t$ the fully loaded unlevered after-tax operating free-cash-flow margin, $r$ a nominal discount rate, $g<r$ continuing nominal growth, and $q$ a terminal-value multiplier. 
Revenues follow a geometric path from $R_0$ to $R_T$, and reference margins rise linearly from $m_0$ to $m_T$. 
We solve:

\[
\lambda P
=\sum_{t=1}^{T}\frac{R_0(R_T/R_0)^{t/T}m_t}{(1+r)^t}
+q\frac{R_Tm_T(1+g)}{(r-g)(1+r)^T},
\qquad m_t=m_0+(m_T-m_0)t/T.
\]

The terminal value represents cash flows beyond the ten-year horizon. 
It assumes a durable commercial position and investment consistent with that margin, including ongoing research, model training, and capacity investment. 
These expenditures are represented through the cash margin rather than forecast separately.

Only the announced company prices and the reported starting sales paces come from observations. We must therefore make assumptions on time horizon, discount rate, cash margin
path, terminal persistence, and operating-value factor. 

\subsection{Results and their main drivers}\label{sec:results}

The reference calculation requires year-ten revenue of \$1.321 trillion for OpenAI and \$1.469 trillion for Anthropic, generating about \$330 billion and \$367 billion of annual free cash flow. 
This is based on a 15\% discount rate, 2\% continuing growth, and cash margins rising from -10\% to 25\%, with $\lambda=q=1$.
From the reported starting proxies, the implied annual average revenue growth rates are 49\% and 41\% respectively. 
If we instead assume a 10\% discount rate and 40\% mature margin, the revenue requirements decrease to \$347 billion and \$386 billion, with annual growth of 31\% and 23\%. 

\begin{figure}
\centering
\includegraphics[width=1\linewidth,height=\textheight,keepaspectratio,alt={Sales growth and operating value under two financial settings. Panels A and B express modeled value as a percentage of the respective financing price (\$852 billion and \$965 billion), starting from annualized sales proxies of \$24 billion and the \$47 billion floor. Markers show the growth rates supporting those hypothetical operating targets and report year-ten revenue. Panel C compares the joint requirements with a hypothetical \$5 trillion annual 2036 final-payer AI software and services pool, including completed work. All cases use ten years of geometric revenue growth, linear margins from -10\% to the stated endpoint, and 2\% continuing growth. The full financing price supplies each operating target (the full financing price); the continuing-value term is included without reduction (no reduction). Panel C assumes no interprovider sales or outside-pool receipts. Neither growth nor the pool is independently forecast; no product mix is specified.}]{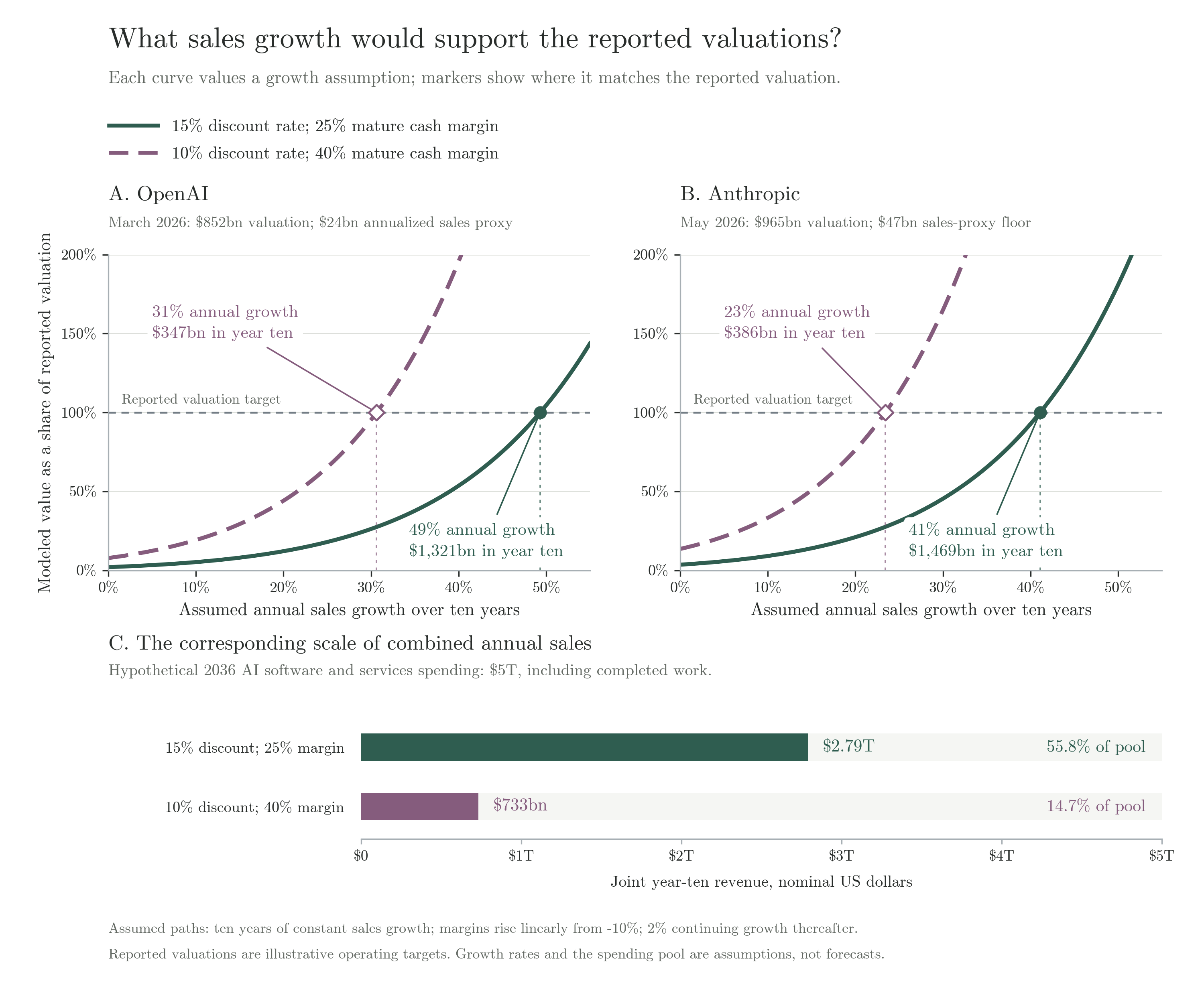}
\caption{Sales growth and operating value under two financial settings. Panels A and B
express modeled value as a percentage of the respective financing price (\$852
billion and \$965 billion), starting from annualized sales proxies of \$24
billion and the \$47 billion floor. Dots mark the growth rates supporting
those hypothetical operating targets and report year-ten revenue. Panel C
compares the joint requirements with a hypothetical \$5 trillion annual 2036
final-payer AI software and services pool, including completed work. All cases
use ten years of geometric revenue growth, linear margins from -10\% to the
stated endpoint, and 2\% continuing growth. The full financing price supplies
each operating target ($\lambda=1$); the continuing-value term is included
without reduction ($q=1$). Panel C assumes no interprovider sales or
outside-pool receipts. Neither growth nor the pool is independently forecast;
no product mix is specified.}\label{fig:access-sales}
\end{figure}

Figure~\ref{fig:access-sales} plots our model of operating value against sales
growth. 
The intersection with the 100\% financing-price line identifies the growth rates necessary to achieve the target valuation. 
If all required revenue comes from access and subscriptions, the 40\% cash margin requires covering serving costs and reinvestment while retaining that share of revenue, suggesting substantial operating leverage and pricing power.

\FloatBarrier

Table~\ref{tbl:valuation-cases} compares the reference and
low-discount/high-margin cases with three further variations: delayed
cash generation, reduced terminal persistence, and a lower
operating-value target. 
The low-discount/high-margin case combines two changes, and the discount rate is the larger lever. 
Lowering the rate to 10\% while keeping reference margins reduces the year-ten endpoints for OpenAI and Anthropic from \$1.321 trillion and \$1.469 trillion to \$571 billion and \$638 billion. 
Raising the mature margin to 40\% at a 15\% discount rate reduces them to \$797 billion and \$879 billion. 
Because margins rise linearly toward the mature level, this change also raises interim margins. 
Appendix~\ref{app:valuation} specifies the schedules and further sensitivities.

\begin{table}[htbp]
\centering
\captionsetup{labelfont=bf,textfont=bf,justification=raggedright,
  singlelinecheck=false,skip=8pt}
\renewcommand{\arraystretch}{1.12}
\begin{threeparttable}
\caption{Required revenue in year ten}\label{tbl:valuation-cases}
\begin{tabular}{@{}
  >{\raggedright\arraybackslash}p{\dimexpr 0.3000\linewidth - 1.8000\tabcolsep\relax}
  >{\raggedright\arraybackslash}p{\dimexpr 0.3800\linewidth - 2.2800\tabcolsep\relax}
  >{\raggedleft\arraybackslash}p{\dimexpr 0.1600\linewidth - 0.9600\tabcolsep\relax}
  >{\raggedleft\arraybackslash}p{\dimexpr 0.1600\linewidth - 0.9600\tabcolsep\relax}@{}}
\toprule
\textbf{Case} & \textbf{Change from reference} & \textbf{OpenAI} &
\textbf{Anthropic} \\
\midrule
Reference & 15\% discount; 25\% mature FCF margin & \$1.321T &
\$1.469T \\
Low discount, high margin & 10\% discount; 40\% mature margin & \$0.347T
& \$0.386T \\
Delayed cash generation & Negative FCF in years 1-5; zero in year 6 &
\$1.440T & \$1.634T \\
Half terminal contribution & \(q=0.5\) instead of 1 & \$2.168T &
\$2.389T \\
Lower hypothetical operating target & \(\lambda=0.75\) instead of 1 &
\$0.982T & \$1.090T \\
\bottomrule
\end{tabular}
\begin{tablenotes}[flushleft]
\footnotesize
\setlength{\itemsep}{3pt}
\item[] \textit{Notes.} Values are trillions of nominal US dollars, rounded
to three decimals. Hypothetical operating-value targets are based on
OpenAI's \$852 billion and Anthropic's \$965 billion financing prices.
\item[] \textit{Reference assumptions.} All cases use ten years and the
reported starting sales proxies. The reference uses 15\% discounting,
margins rising linearly from -10\% to 25\%, 2\% continuing growth,
the full financing price as target ($\lambda=1$), and an unscaled
continuing-value term ($q=1$). The delayed margin schedule is in Appendix~\ref{app:valuation}.
\end{tablenotes}
\end{threeparttable}
\end{table}

\FloatBarrier

Durable, long-term cash generation is necessary in our scenarios to justify the valuations. 
Roughly three quarters of the calculated value (75\% for OpenAI and 74\% for Anthropic), comes from cash flows after year ten. 
So these scenarios assume substantial growth followed by years of sustained cash generation. 
If we want a more conservative approach, we could for example halve the terminal contribution, but then we need higher growth in revenue before that, and the effect is even greater than early-on delays in positive cash generation, increasing year ten revenue by about 63\%.
Setting continuing growth past year ten to zero raises the combined year-ten revenue requirement from \$2.79 trillion to \$3.16 trillion in the reference case and from \$733 to \$892 billion in the low-discount/high-margin case (Table~\ref{tbl:terminal-growth}).

Early losses also make the case more strenuous. %
The reference OpenAI path generates a net \$14.6 billion over its first five modeled years.
A September 2026 \emph{Financial Times} report of a private presentation instead describes
a projected \$278 billion free-cash-flow deficit during 2026-2030
\citep{ft2026openai}. 
As a synthetic stress test, replacing the first five model-year cash flows
with evenly allocated losses totaling that deficit raises OpenAI's year-ten
requirement to \$1.632 trillion; the model periods approximate rather than
reproduce the reported calendar years.\footnote{The report postdates the financing, so may not have been figured in by investors. For details on timing and our assumptions on the annual allocation of losses, see Appendix~\ref{app:valuation}.}

Anthropic's disclosures make sustaining the assumed cash margins a more
demanding condition of the access-sales case.
In 2025, operating losses exceeded \$8 billion on nearly \$4.6 billion of
revenue \citep{reuters2026anthropicfinancials}; later reports of improved
earnings do not yet establish the cash margins used here.%
\footnote{\interlinepenalty=10000 Reuters relayed a September report that Anthropic expected positive
adjusted operating income for a second consecutive quarter, without a
reconciliation to free cash flow \citep{reuters2026anthropicadjustedprofit}.
The roughly \$42 billion 2025 net loss included a roughly \$34 billion
noncash financing-related accounting charge
\citep{reuters2026anthropicfinancials}. These measures concern different
periods and should not be read as annual operating cash deficits.}
Reuters also reported at least \$518 billion of infrastructure obligations
over about a decade, approximately 80\% noncancelable or payable regardless
of usage \citep{reuters2026anthropiccommitments}.
Securing capacity can enable growth when compute constrains paying demand,
but less flexible payments leave Anthropic exposed if customers' spending
falls short; many large customers reportedly lack long-term contracts
\citep{reuters2026anthropicdistribution}.
The financial challenge is therefore to fill committed capacity at prices
that cover its costs while funding further model development.
These disclosures give reason to scrutinize the margin path as well as
sales growth: weaker sustained cash margins would require still higher
revenue to support the same valuation. Appendix~\ref{app:anthropic-financial-evidence}
sets out the evidence and its limits.

Variations in the initial sales estimates have smaller impacts on the requirements than the above considerations. 
Doubling the sales figures while holding the margin paths and other settings fixed lowers the reference endpoints by only 2-3\%.

Later valuation discussions change the scale of the situation however. 
Table~\ref{tbl:higher-valuation} varies only the hypothetical operating targets, holding the \$24 billion and \$47 billion starting proxies, original ten-year horizon, margin paths, and terminal assumptions fixed. 
At \$1.43 trillion for OpenAI and \$2 trillion for Anthropic, combined year-ten revenue rises to \$5.372 trillion in the reference setting or \$1.413 trillion at a 10\% discount rate and 40\% mature margin. 

\subsection{Could access and seats supply the necessary revenue?}\label{sec:access-sales}

We next examine whether revenue from model access alone could support
the announced valuations, through growth in the number of customers and the intensive margin (higher spending per customer, via greater API usage or higher-priced subscription plans).

OpenAI reports roughly fourfold output-token growth from June 2025 to March 2026 among firms already using ChatGPT Enterprise by June 2025 \citep[sec.~4.1]{chatterji2026organizations}.
The measure covers ChatGPT Enterprise subscriptions, not separately billed API revenue.  Scientific research is another potential source of model demand, illustrated by OpenAI's deployment agreement with the U.S. National Laboratories \citep{openai2025nationallabs}. 
Investor accounts likewise emphasize enterprise adoption and complex workflows involving agents \citep{bessemer2024,anthropic2026g,anthropic2026h,lightspeed2025}. 
Usage of this kind grows only as customers integrate models into their
systems and redesign workflows around them, and providers increasingly
pay for part of that work. Anthropic has committed \$100 million to train
10,000 forward-deployed engineers, who build deployments with customers,
by the end of 2027 \citep{anthropic2026academy}, presumably hoping to recover some of this investment through increased customer spending on access. 

If paying demand is constrained by available compute, capacity expansion
provides a further basis for expecting rapid access-sales growth. 
Applying OpenAI's reported revenue-to-capacity ratio to Epoch's reports of data center capacity plans illustrates the possibility of roughly \$100~billion in annualized revenue by the end of 2028 \citep{friar2026business,epoch2026datacenters}.\footnote{OpenAI reported
annualized revenue exceeding \$20~billion and approximately 1.9~GW of
compute in 2025. Following Epoch's interpretation of these figures as IT
capacity \citep{epoch2026chipusersmethodology}, its October~5, 2026 dataset
projects 7.95~GW across eight US and UAE Stargate sites by the end of 2028.
Holding other 2025 capacity constant and subtracting the 0.21~GW already
included gives $1.9-0.21+7.95=9.64$~GW, hence approximately
$20(9.64/1.9)\simeq100$~billion dollars in annualized revenue, or about
72\% annual growth over three years. This naturally assumes project
completion, sufficient paying demand, and unchanged revenue per GW.}

Anthropic's distribution arrangements illustrate why expanding access
sales need not deliver proportionate growth in cash available to the firm.
In 2025, 47\% of its revenue passed through cloud marketplaces; distribution
fees absorbed about 16\% of those sales \citep{reuters2026anthropicdistribution}.%
\footnote{\interlinepenalty=10000 The reported amounts were roughly \$2.16 billion of marketplace sales and \$351 million of fees, equivalent to about 7.6\% of total 2025 revenue.  Anthropic recognized these contracts gross and recorded the fees as operating expenses \citep{reuters2026anthropicdistribution}. The fully loaded cash margin must cover these fees once; the historical fee share is not projected forward.}
Cloud partners help reach customers, but also receive payments for computing capacity and may serve and develop competing models \citep{reuters2026anthropiccommitments}. 
The valuation case consequently depends on frontier model providers retaining durable
commercial positions as this market expands.

Rapid expansion in model capabilities, such as through recursive self-improvement, could also expand these markets while reducing development and inference costs.
OpenAI has stated as a business goal the creation of an automated AI researcher \citep{altman2026plan,pachocki2026alien}.
Anthropic also reports using its models in development, with substantial portions of their code written by their own models, and takes seriously the prospect that AI systems could increasingly autonomously develop their own successors \citep{favaro2026aibuilds}.

The idea is not new - Sequoia's Dan Roberts discussed compounding investment in intelligence
in 2023 \citep{roberts2023intelligence}. 
The financial effect still depends on customer payments, costs, and competition however.
\footnote{OpenAI discussed AI researchers and RSI before the financing round in question \citep{openai2025progress}. 
The June and September priority statements cited here, together with Anthropic's September update, provide later public statements about this strategy.}

An illustrative allocation shows the required scale of sales growth. 
OpenAI reported more than 50 million consumer subscribers and 9 million paying business
users in February 2026 \citep{openai2026scale}. 
Given our price and mix benchmarks (Appendix~\ref{app:allocation}, which also includes alternative mixes), 200 million consumer
subscriptions and 100 million business seats would generate \$52.8 billion
annually, using the \$240 annual business-seat benchmark.
The remaining separately billed API revenue would need to reach about \$295 billion in the low-discount/high-margin case or \$1.269 trillion in the reference case.

Competition from open-weight models could put pressure on prices.
But access to model weights does not remove the need for computing
infrastructure, non-public training data, and the systems needed to
serve customers, which can still make entry difficult \citep{azoulay2024}.
Open-weight model releases are also not guaranteed to continue, especially if the firms behind them achieve closer to parity with the frontier. 

Some technical improvements that improve the user experience can actually harm revenue. 
Cache billing, model routing, and more efficient tools can reduce spending
\citep{openrouter2026caching,wendell2026codingcosts}. 
Short-run price responses across providers of the same open-source models show that some customers are savvy enough to switch among providers \citep[sec.~7]{demirer2025intelligence}. 
As one illustration, the cloud-platform provider Vercel released data on usage through their platform indicating that OpenAI nearly doubled token volume from June to July 2026 but only achieved about 15\% growth valued at published prices \citep{vercel2026july,vercel2026august}. 

\subsection{Spending capture and the cost of delivery}\label{sec:spending-capture}

We next ask what share of future AI customer spending OpenAI and Anthropic
would need to capture to support their announced valuations, under different cash-margin assumptions. 
Imagine that in 2036 individuals and businesses ultimately spend some amount $S_T$ on AI-enabled software and services, including completed work, across all providers.
We consider hypothetical annual totals of \$2 trillion, \$5 trillion, and \$10 trillion to show how the results depend on the assumed spending pool. 
The two firms' revenue from this global pool depends on two shares: the fraction $c$ of final-customer spending that reaches model providers, and the fraction $s$ of those
providers' receipts that goes to OpenAI and Anthropic. 
After eliminating interprovider sales, their joint revenue is
\[
R^{OA}_T = S_T c s + R^{OA}_{\mathrm{other},T},
\]

where $R^{OA}_{\mathrm{other},T}$ denotes revenue from outside the software-and-services pool, such as advertising.
Here we ask whether customer spending on AI software and services could meet the calculated revenue requirements, so we exclude advertising and the other possible revenue streams discussed in Section~\ref{sec:commercial-routes}.

\begin{figure}
\centering
\includegraphics[width=1\linewidth,height=\textheight,keepaspectratio,alt={Three panels show the joint share of hypothetical annual 2036 spending needed by OpenAI and Anthropic. Pools are \$2 trillion, \$5 trillion, and \$10 trillion, on common axes. At a 25\% mature cash margin and 15\% discount rate, required shares are 139.5\%, 55.8\%, and 27.9\%. At a 40\% margin and 10\% discount rate, they are 36.6\%, 14.7\%, and 7.3\%. The 100\% line marks the assumed pool's size; shares below it still require economic justification.}]{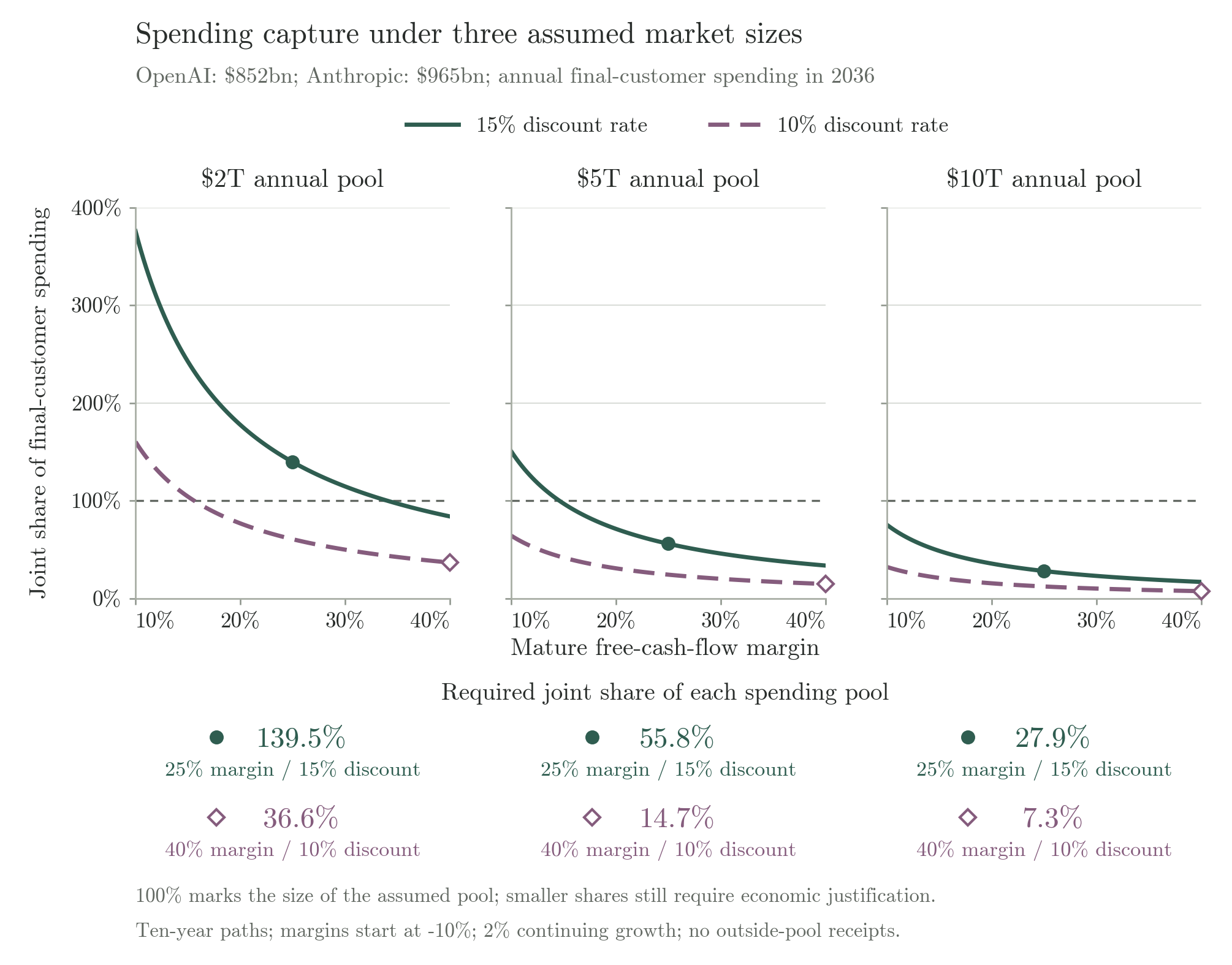}
\caption{Required joint share of hypothetical worldwide annual AI software and services spending in 2036. The panels assume \$2 trillion, \$5 trillion, and \$10 trillion, from left to right, and use common axes. The targets are OpenAI's \$852 billion and Anthropic's \$965 billion financing prices. Solid curves use a 15\% discount rate; dashed curves use 10\%. Circles mark a 25\% mature cash margin at 15\% discounting; diamonds mark a 40\% margin at 10\%. The percentages beneath each panel give the required joint spending share at these points. All cases use ten-year revenue paths, margins rising linearly from -10\%, and 2\% continuing growth. Spending is counted once at the final customer and includes completed work. Interprovider sales and outside-pool revenue are excluded. The spending pools are assumed.}\label{fig:feasibility}
\end{figure}

Panel C of Figure~\ref{fig:access-sales} retains the \$5 trillion illustration, 
while Figure~\ref{fig:feasibility} compares annual spending pools of \$2 trillion, \$5 trillion, and \$10 trillion. 
Both figures set outside-pool revenue to zero.
Under a 15\% discount rate and 25\% mature cash margin, the combined \$2.79 trillion 
revenue requirement implies shares of 139.5\%, 55.8\%, and 27.9\%, respectively. 
Under a 10\% discount rate and 40\% mature margin, the \$733 billion requirement 
implies shares of 36.6\%, 14.7\%, and 7.3\%. 
These shares are $cs$, the fraction of total final-customer spending
that must reach OpenAI and Anthropic together.

A share above 100\% means the assumed pool cannot supply the required revenue. 
A share below 100\% leaves open whether the firms can achieve the required spending share 
and cash margin together.
Moving into service delivery could raise the model provider's share of customer spending while lowering its cash margin, so both effects must be assessed together.
Appendix~\ref{app:spending} reports further financial settings. 

Provider pricing also affects the downstream adoption and competitive conditions
behind these spending shares. 
Usage fees raise adopters' production costs, while access fees can make adoption or market entry unprofitable \citep{athey2025}. 
Higher fees can thus increase receipts while also narrowing future revenue possibilities.

\FloatBarrier

\section{Commercial routes beyond model access}\label{sec:commercial-routes}

\begin{figure}[htbp]
\centering
  \includegraphics[width=1\linewidth,height=\textheight,keepaspectratio,alt={Selected productive activities and financial claims. Solid arrows show disclosed relationships; dotted arrows mark undisclosed provider financial links. The appendix table provides a wider comparison.}]{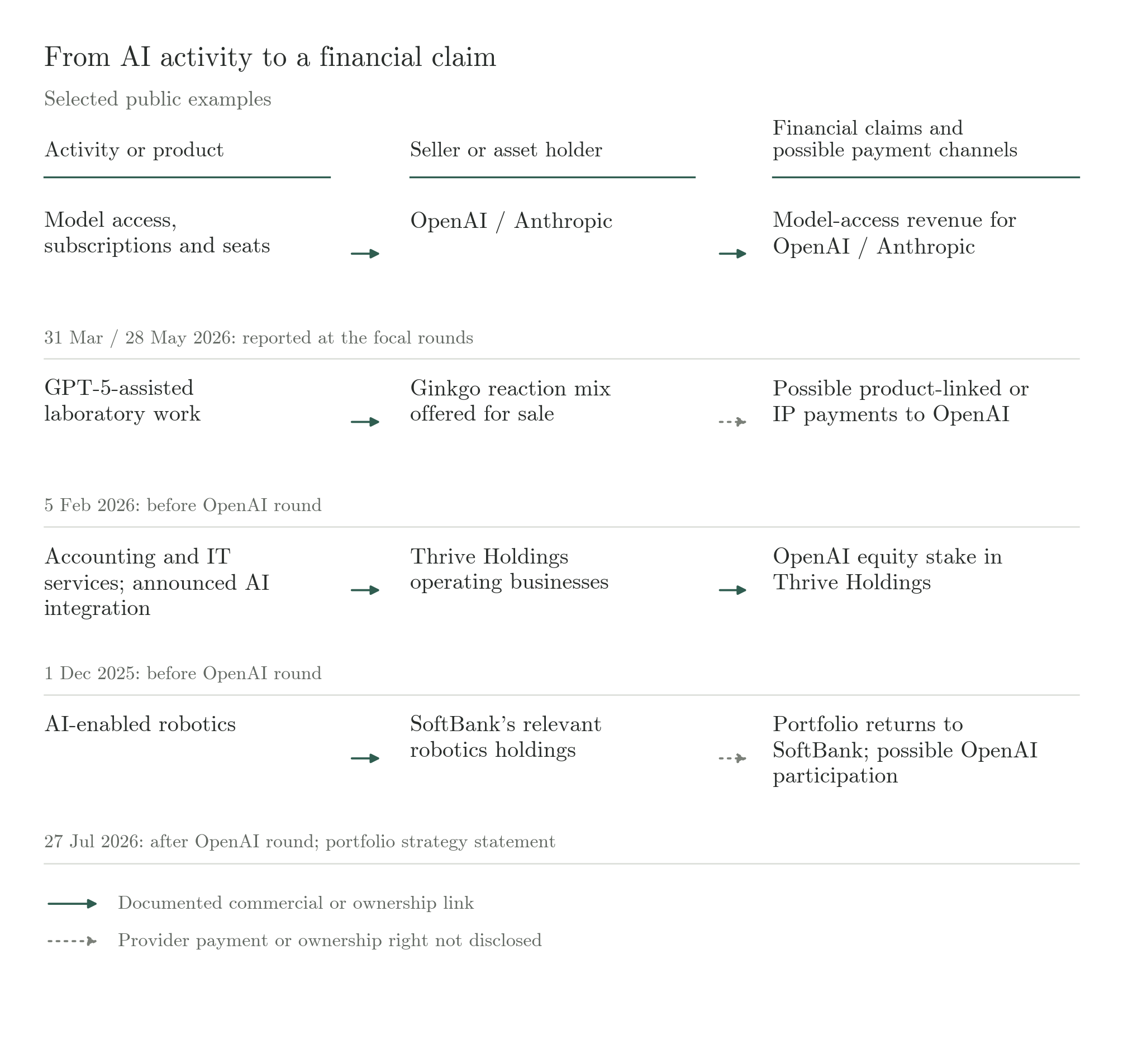}
\caption{Selected productive activities and financial claims. Solid arrows show
disclosed relationships; dotted arrows mark undisclosed provider financial
links. Ginkgo's product, OpenAI's equity stake in Thrive Holdings, and SoftBank's holdings
establish different revenue claims \citep{openai2026ginkgo,ginkgo2026lab,openai2025thrive,softbank2026}.
Table~\ref{tbl:commercial-rights} gives the wider comparison.}\label{fig:return-rights}
\end{figure}

Even where hard revenue figures are difficult to pin down, we have reason to believe frontier model companies may be pursuing additional sources of income or financial value, such as through advertiser payments, application-linked payments, and ownership interests in businesses that use AI. 
If (scarce) frontier AI is a key input to developing new pharmaceutial products, for example, it would make sense for model providers to try to capture some of those gains beyond the mere price of the tokens involved. 
Figure~\ref{fig:return-rights} distinguishes the productive activity, its seller or asset holder, and the frontier provider's financial claim. 
Table~\ref{tbl:commercial-rights} records the arrangements and their timing.

These routes contribute to valuation in different ways. Sales and contractual
payments enter operating cash flows. For controlled service businesses, the
parent's claim depends on cash flows attributable to it after outside
investors' claims. A minority holding contributes investment value and
possible distributions or realized gains, rather than the investee's gross
revenue. A combined valuation would need to distinguish these claims and
avoid counting the same underlying cash flows twice.

In March, OpenAI reported that its advertising pilot had reached more than \$100 million in annual recurring revenue in under six weeks \citep{openai2026funding}. 
OpenAI's CFO also proposed licensing, intellectual-property agreements, and outcome pricing for scientific applications on January 18, before the March financing round \citep{friar2026business}. 
A third route is through ownership of service businesses. 
OpenAI announced a stake in Thrive Holdings in December 2025 \citep{openai2025thrive}. 
In May 2026 it announced majority ownership and control of the separate OpenAI Deployment Company \citep{openai2026deployment}.
Axios reported a minimum 17.5\% investor return and capped profits
\citep{axios2026deploymentterms}; an earlier report described five-year
exits with an OpenAI guarantee \citep{axios2026servicesventures}.
These reported preferences and obligations qualify OpenAI's residual claim,
but their net financial effect cannot be quantified from the available terms.
We do not interpret the percentage as an annual IRR or use it as an
operating-cost adjustment.

Accenture gives an idea of the scale such service businesses can reach. 
Its fiscal 2025 cash from operations less property-and-equipment purchases was
\$10.9 billion \citep{accenture2025}.
The annual cash flows at the end of our two reference paths, \$330 billion and \$367 billion, are about 30 and 34 times that amount. A services business of Accenture's size would
therefore cover roughly 3 percent of the required cash flow, unless
AI-enabled services prove far more productive or profitable. %

Scientific work raises similar payment and ownership questions for discoveries and products. 
In OpenAI's collaboration with Ginkgo, GPT-5 proposed protein-synthesis experiments.
Ginkgo's automated laboratory then tested them, and the partners reported a
40\% production-cost reduction for one protein system \citep{openai2026ginkgo}. 
Ginkgo then announced a reaction-mix product derived from the work \citep{ginkgo2026lab}. 
Ginkgo sells the product; any royalty or ownership share for OpenAI
would depend on terms that have not been made public.

Anthropic has also brought scientific capabilities inside the firm. 
An April report described its acquisition of Coefficient Bio \citep{techcrunch2026coefficient}. 
The work includes teams focusing on enzyme discovery and drug-discovery, 
as well as a Genentech proof of concept linking agents to a partner's laboratory
instruments \citep{anthropic2026enzyme,anthropic2026mhs}. 

\Needspace{5\baselineskip}
Comparisons with other firms illustrate alternative arrangements for ownership, payment, and development responsibilities. 
Isomorphic Labs, which was spun out of Google DeepMind, has an internal pipeline of drug candidates \citep{isomorphic2026}. 
This could generate revenue from product sales, but also incurs costs for testing, regulatory approval, development, and commercialization \citep{agrawal2024discovery}.

LLMs, despite their initial focus on language, have also come to be seen as useful for robotics. 
This could create complementarities between SoftBank's robotics holdings and its model provider investments \citep{softbank2025,softbank2026,softbank2026followon}, though we found no disclosed revenue channels to that effect.

The boundary between AI providers and application firms depends on their relative capabilities and the costs of coordinating and adapting complementary activities \citep{jacobideswinter2005}. 
Ginkgo linked GPT-5's experimental design capabilities to a defined laboratory workflow, allowing model provision and laboratory execution to remain separate \citep{openai2026ginkgo}.
Figure, by contrast, described a mismatch between OpenAI's broad remit and its own need for robot-specific model development \citep{techcrunch2025figure}.
Figure's decision suggests that when AI requires continual adaptation to a firm's physical product, an internal team may be better able to align development priorities and coordinate changes, despite the costs of building those capabilities \citep{williamson1991,jacobideswinter2005}. 
The contrast therefore concerns whether the application can use externally supplied AI
through a defined interface or requires control over its continuing development. 
For frontier providers, this creates a limit on value capture: an application firm may benefit from AI while internalizing the capabilities that differentiate its product, leaving the provider's returns dependent on its remaining contribution and financial rights \citep{teece1986}.

\section{Conclusion}\label{sec:conclusion}

We find that, at least under certain demanding assumptions, revenue from access and business seats could generate cash flows consistent with the selected valuation targets. 
With the full financing prices used as hypothetical operating-value targets, combined annual revenue would need to reach \$2.79 trillion by year ten in the reference case. 
Under the more optimistic combination of a lower discount rate and a higher mature margin, the requirement falls to \$733 billion. 
These estimates are naturally highly sensitive to the discount rate, cash margins, and the value of earnings beyond the ten-year horizon. 

Under the higher valuation target identified in recent news sources, the corresponding combined revenue requirements rise to \$5.372 trillion and \$1.413 trillion. 
Public documents also identify other potential sources of value, including application-specific payments, equity stakes in service businesses, and scientific collaborations. 

We identify a tension in the valuation numbers that points to the possible relevance of other sources of income or financial value, although we cannot establish how investors account for them.
Greater clarity on these business models would help assess whether frontier AI firms' valuations are justified by their revenue prospects or reflect irrational exuberance. 
It could also inform research on the broader effects of AI on scientific innovation and labor markets. 
In particular, a clearer picture of revenue sources, from advertising to scientific partnerships, could guide research on how the associated applications complement or substitute for human labor across tasks and jobs, and how they affect productivity.

\section{Disclosures}

\noindent \textbf{Data availability}: Code and data to replicate the analysis and figures is available at an anonymized repo at \url{https://anonymous.4open.science/r/AI-valuation-paper/}.\\

\noindent \textbf{Conflict of Interest}: The authors declare no conflict of interest.\\

\noindent \textbf{Declaration of Generative AI Use}: The authors used GPT-6 (OpenAI) and Claude Sonnet-5.5 (Anthropic) for source discovery and organization; assistance with code development, including tests, numerical checks, and figure production; and for critical review and editing. The authors defined the research question, directed the analysis, wrote the manuscript, and reviewed and verified AI-assisted outputs. The authors take full responsibility for the paper's content, analyses, claims, and conclusions.

\bibliography{references}
\addcontentsline{toc}{section}{References}

\manuscriptappendix

\section{Data and methods}\label{app:methods}

\subsection{Documentary selection, classification, and timing}
\label{app:selection}

We collected 61 public documents to examine evidence of business model clues and valuations. This includes official OpenAI and Anthropic announcements from January 2025 through October 6, 2026 concerning financing, ventures, equity interests, controlled businesses, and payment or distribution rights.
We also include investor statements, transaction reports, pricing records, and scientific collaborations, but did not specifically look for routine customer and partnership agreements. 

From these, we had Codex code the speaker, proposed mechanism,
beneficiary, evidentiary status, and timing.
AI-assisted passage checks are recorded for each source; subsequent author source review is tracked separately.
Table~\ref{tbl:commercial-rights} records the arrangements and their timing relative to the disclosure of investment rounds. 

\Needspace{14\baselineskip}

\begingroup
\small

\begin{longtable}[]{@{}
  >{\raggedright\arraybackslash}p{\dimexpr 0.34\linewidth - 1.36\tabcolsep\relax}
  >{\raggedright\arraybackslash}p{\dimexpr 0.48\linewidth - 1.92\tabcolsep\relax}
  >{\raggedright\arraybackslash}p{\dimexpr 0.18\linewidth - 0.72\tabcolsep\relax}@{}}
\caption{Commercial arrangements and disclosed financial claims}\standardlabel{tbl:commercial-rights}\tabularnewline
\toprule\noalign{}
\begin{minipage}[b]{\linewidth}\raggedright
Activity and possible billing
\end{minipage} & \begin{minipage}[b]{\linewidth}\raggedright
Financial claim or role
\end{minipage} & \begin{minipage}[b]{\linewidth}\raggedright
Timing relative to round
\end{minipage} \\
\midrule\noalign{}
\endfirsthead
\toprule\noalign{}
\begin{minipage}[b]{\linewidth}\raggedright
Activity and possible billing
\end{minipage} & \begin{minipage}[b]{\linewidth}\raggedright
Financial claim or role
\end{minipage} & \begin{minipage}[b]{\linewidth}\raggedright
Timing relative to round
\end{minipage} \\
\midrule\noalign{}
\endhead
\bottomrule\noalign{}
\endlastfoot
AI assistance or task completion sold through subscriptions, usage, or
integrated products \citep{openai2026funding,anthropic2026h}. &
Disclosed provider sales: customer payments can become model-company
revenue; realized product margin is undisclosed. & Sales paces disclosed
with the rounds. \\
\addlinespace[3pt]
Consumer attention sold to advertisers \citep{openai2026funding}. &
Disclosed provider monetization: OpenAI reported a pilot above \$100
million ARR; an advertiser pays, rather than the software user. & With
OpenAI's March round. \\
\addlinespace[3pt]
AI-assisted laboratory experiments and a resulting reaction mix sold by
Ginkgo \citep{openai2026ginkgo,ginkgo2026lab}. & Product sales accrue
to Ginkgo; OpenAI's contractual entitlement, if any, is undisclosed. &
Before OpenAI's March round. \\
\addlinespace[3pt]
Accounting and IT services with announced AI integration in Thrive Holdings businesses
\citep{openai2025thrive}. & Disclosed equity right: OpenAI announced
a stake in Thrive Holdings; its share and attributable profit are
undisclosed. Customer service revenue belongs to the operating seller. &
Before OpenAI's March round. \\
\addlinespace[3pt]
Biotechnology AI research capacity acquired through Coefficient Bio
\citep{techcrunch2026coefficient}. & Reported Anthropic acquisition;
no drug candidate, medicine sales, or external royalty is identified. &
Before Anthropic's May round. \\
\addlinespace[3pt]
Mid-market implementation and custom systems through a new services
company \citep{anthropic2026services}. & Reported equity participation:
transaction reporting identifies a minority stake and a roughly \$300 million
investment commitment \citep{axios2026servicesventures,techcrunch2026servicesventures}.
The official release leaves Anthropic's interest unspecified; its precise
ownership share, contractual rights, and attributable returns remain
unverified. & Before Anthropic's May round. \\
\addlinespace[3pt]
Implementation and workflow redesign by OpenAI's deployment company
\citep{openai2026deployment}. & Disclosed control right: OpenAI
announced majority ownership of the deployment firm. Its attributable
returns depend on delivery costs and outside investors' terms
(Section~\ref{sec:commercial-routes}). &
After OpenAI's March round. \\
\addlinespace[3pt]
Enterprise AI sales through the planned Japanese venture
\citep{softbank2025jv}. & Announced 50:50 venture consolidated by
SoftBank Corp.; OpenAI's possible venture return differs from gross
venture revenue. & Before OpenAI's March round. \\
\addlinespace[3pt]
Data-center development and leasing by SB Energy
\citep{openai2026sbenergy}. & OpenAI and SoftBank each announced
\$500 million of investment; any OpenAI investment return and its lease
payments are separate flows. & Before OpenAI's March round. \\
\addlinespace[4pt]
\multicolumn{3}{@{}l}{\textit{Materials-science comparisons}} \\*
\addlinespace[2pt]
An equity interest in a materials-technology business held by XtalPi
\citep[419, 434]{xtalpi2024prospectus}. & Comparison: XtalPi reported
30\% of Xinshengtai (Hangzhou) Materials Technology as of May 29, 2024;
control and attributable product earnings are not established. &
June 2024 disclosure. \\*
\addlinespace[3pt]
Joint AI, robotics, and new-energy-materials R\&D by XtalPi and GCL
\citep[17]{xtalpi2024interim}. & Expected R\&D service fees to XtalPi
above US\$135 million over five years. & August 2024 agreement. \\*
\addlinespace[3pt]
AI design and physical validation through CuspAI's materials foundry and
A*STAR laboratories \citep{cuspai2026foundry}. & Comparison:
laboratory partnership disclosed; equity, discovery-IP allocation, and
rights to resulting materials sales are unspecified. & July 2026;
after both rounds. \\
\end{longtable}

\noindent\begin{minipage}{\linewidth}
\footnotesize
\textit{Notes.} The XtalPi and CuspAI rows illustrate possible financial arrangements with an AI company but do not involve Anthropic or OpenAI.
Source passages and dates are recorded in the replication data.
\end{minipage}

\endgroup

\subsection{Reverse-valuation specification and
sensitivities}\label{app:valuation}

We calculated the annual revenue needed to support the two firms' announced
valuations under alternative financial assumptions (Section~\ref{sec:financial-conditions}).
In the reference paths, margins become positive in year three; 
the present values of negative interim cash flows are approximately \$3.2 billion for OpenAI and \$5.9 billion for Anthropic. 
The delayed margins for years one to ten are
-10\%, -10\%, -8\%, -6\%, -4\%, 0\%, 5\%, 12\%, 19\%, and 25\%;
the corresponding present values of negative cash flows are about
\$20 billion and \$32 billion.
The half-terminal and lower-target cases change only $q$ to 0.5 and $\lambda$ to 0.75, respectively.
A five-year horizon under reference rates and
margins gives \$727 billion and \$815 billion revenue endpoints but
requires 98\% and 77\% annual growth. 
Doubling the starting proxies lowers the ten-year reference endpoints by only 2-3\%. 
The replication package supplies all 648 combinations of the specified horizons,
discount rates, terminal growth rates, initial and mature margins, and
operating-value factors. 

\Needspace{12\baselineskip}

\begin{table}[htbp]
\centering
\caption{Required year-ten revenue under alternative continuing growth rates}\label{tbl:terminal-growth}
\begin{tabular}{@{}p{0.48\linewidth}rr@{}}
\toprule
Financial setting & 2\% growth & Zero growth\\
\midrule

Reference: 15\% discount, 25\% mature margin & \$2.791T & \$3.157T\\
Optimistic combination: 10\% discount, 40\% mature margin & \$0.733T & \$0.892T\\
\bottomrule
\end{tabular}
\noindent\begin{minipage}{\linewidth}
\footnotesize
\textit{Notes.} Combined OpenAI and Anthropic revenue in trillions of
nominal US dollars. Each row changes only continuing growth, retaining
its discount rate, margin path, ten-year horizon, starting sales proxies,
and $\lambda=q=1$. Zero growth still assumes that the mature cash flow
continues indefinitely.
\end{minipage}\par\medskip
\end{table}

We also conduct a sensitivity analysis in which we replace OpenAI's first five modeled
annual cash flows with a \$278 billion deficit reported for calendar 2026-2030
\citep{ft2026openai}. 
An even allocation is -\$55.6 billion annually. %
If so, holding other assumptions fixed, this requires a \$1.632 trillion endpoint;
allocating all the deficit to year one or five requires \$1.722 trillion
or \$1.554 trillion. 
These are the largest and smallest discounted
deficits only under nonpositive annual flows summing to the reported total. 

The model periods approximate, rather than reproduce, the reported calendar
years. %
The Financial Times article's \$856 billion compute-spending forecast is not added to the free cash flow deficit. 
Its \$350 billion 2030 revenue projection does not specify the annual cash path or channel mix.

Applying the same evenly allocated first-five-year loss stress with a
10\% discount rate and 40\% mature margin raises OpenAI's required
year-ten revenue from \$347 billion to \$443 billion. 
This changes its early cash flows without changing the later revenue and margin rules;
no corresponding loss forecast is imposed on Anthropic.

We look at ServiceNow and Accenture as benchmarks to illustrate subscription and service economics. 
Cash from operations less property-and-equipment purchases equals 34.5\% of fiscal 2025 revenue
for ServiceNow and 15.6\% for Accenture \citep{servicenow2025,accenture2025}. 
Subtracting stock compensation gives 19.7\% and 12.6\%. 
For Accenture, the respective cash amounts are
\$10.9 billion and \$8.8 billion against \$69.7 billion of revenue.
These proxies are not estimates of the model's unlevered after-tax operating margin. 
These are of course simply illustrative, as working-capital timing, interest, acquisitions, and investment needs also affect comparisons across firms.

\phantomsection\label{app:higher-valuation}

Based on recent reports of possible higher valuations, in Table~\ref{tbl:higher-valuation} we consider the sensitivity in terms of the corresponding operating-value targets.

The table compares five target pairs under each financial setting, retaining the original starting-sales proxies and model horizon. 
The replication data also records each firm's cash-flow components, required growth, and
source identifiers.

\begin{table}[htbp]
\centering
\small
\caption{Higher hypothetical operating-value targets: required year-ten revenue}\label{tbl:higher-valuation}
\begin{tabular}{@{}lrrrrrr@{}}
\toprule
Setting & \multicolumn{2}{c}{Targets (\$T)} & \multicolumn{3}{c}{Year-ten revenue (\$T)} & \$5T pool \\
& OpenAI & Anthropic & OpenAI & Anthropic & Joint & Share \\
\midrule
15\%; 25\% & 0.852 & 0.965 & 1.321 & 1.469 & 2.791 & 55.8\% \\
15\%; 25\% & 1.400 & 0.965 & 2.204 & 1.469 & 3.673 & 73.5\% \\
15\%; 25\% & 1.430 & 0.965 & 2.253 & 1.469 & 3.722 & 74.4\% \\
15\%; 25\% & 0.852 & 2.000 & 1.321 & 3.119 & 4.441 & 88.8\% \\
15\%; 25\% & 1.430 & 2.000 & 2.253 & 3.119 & 5.372 & 107.4\% \\
\addlinespace
10\%; 40\% & 0.852 & 0.965 & 0.347 & 0.386 & 0.733 & 14.7\% \\
10\%; 40\% & 1.400 & 0.965 & 0.580 & 0.386 & 0.965 & 19.3\% \\
10\%; 40\% & 1.430 & 0.965 & 0.593 & 0.386 & 0.978 & 19.6\% \\
10\%; 40\% & 0.852 & 2.000 & 0.347 & 0.820 & 1.168 & 23.4\% \\
10\%; 40\% & 1.430 & 2.000 & 0.593 & 0.820 & 1.413 & 28.3\% \\
\bottomrule
\end{tabular}
\par\smallskip
\begin{minipage}{\linewidth}\footnotesize
\textit{Notes.} Settings give the discount rate and mature free-cash-flow margin, only the targets change. Starting annualized sales proxies remain \$24B and \$47B, with the original ten-year geometric revenue paths, linear margins from -10\%, 2\% continuing growth, and the full terminal term ($q=1$). Each price is used at face value as a hypothetical operating target; no measured equity-to-operating-value bridge is available. 
The \$1.4T OpenAI price is a reported pre-money proposal; \$1.43T assumes \$30B of new primary equity. Anthropic's \$2T is an attributed investor IPO expectation, not an agreed offering price. Its reported \$190--200B 2028 revenue forecast supplies context and is not used directly in the model. Shares assume the stated final-payer software-and-services pool, no interprovider sales and zero outside-pool receipts. Above 100\% rules out the combination under this pool, perimeter and financial setting.
\textit{Sources.} Historical calibrations \citep{openai2026funding,anthropic2026h}; OpenAI discussions \citep{bloomberg2026openaitarget,bloomberg2026uaefunds}; Anthropic expectation, forecast and filing \citep{ft2026anthropicipo,reuters2026anthropicforecast,anthropic2026s1,reuters2026anthropicipo}.
\end{minipage}
\end{table}

\FloatBarrier

\subsection{Anthropic financial evidence}\label{app:anthropic-financial-evidence}

Table~\ref{tbl:anthropic-financial-evidence} distinguishes reported historical results, subsequent earnings claims, and future purchase obligations. 
The quarterly earnings item is a Reuters relay of Financial Times reporting that Reuters said it could not immediately verify. 

\begingroup
\small
\setlength{\tabcolsep}{4pt}
\renewcommand{\arraystretch}{1.13}
\begin{longtable}{@{}
>{\raggedright\arraybackslash}p{0.24\linewidth}
>{\raggedright\arraybackslash}p{0.30\linewidth}
>{\raggedright\arraybackslash}p{\dimexpr0.46\linewidth-4\tabcolsep\relax}@{}}
\caption{Anthropic financial measures and limits on their use}\standardlabel{tbl:anthropic-financial-evidence}\tabularnewline
\toprule
Measure & Reported amount or status & Interpretation for the valuation exercise\\
\midrule
\endfirsthead
\multicolumn{3}{@{}l}{\textit{Table~\ref{tbl:anthropic-financial-evidence}, continued}}\\
\toprule
Measure & Reported amount or status & Interpretation for the valuation exercise\\
\midrule
\endhead
\bottomrule
\endlastfoot
Recognized revenue & Nearly \$4.6B, FY2025
\citep{reuters2026anthropicfinancials}. &
Annual historical revenue; different period and measure from the
May 2026 annualized starting proxy.\\
\addlinespace
Operating result & Loss above \$8B; \$12.65B total operating
expenses, FY2025 \citep{reuters2026anthropicfinancials}. &
Accounting operating performance; not an observed free-cash-flow margin.\\
\addlinespace
Compute and infrastructure & \$7.33B, FY2025
\citep{reuters2026anthropicfinancials}. &
Aggregate cost measure, without a training/serving split or a cash-flow
reconciliation.\\
\addlinespace
Net loss & About \$42B, including roughly \$34B of
financing-related accounting charges, FY2025
\citep{reuters2026anthropicfinancials}. &
Noncash financing revaluation must not be imposed as operating cash burn.\\
\addlinespace
Liquidity & \$20.28B of cash, cash equivalents and short-term
investments at December 31, 2025 \citep{reuters2026anthropicfinancials}. &
A stock at an earlier date; not an annual cash flow or a financing-date
net-cash adjustment.\\
\addlinespace
Revenue composition & About \$3.8B consumption-based and
\$789M subscriptions, FY2025 \citep{reuters2026anthropicdistribution}. &
Supports examining paid access; does not establish the modeled API/seat
categories or future mix.\\
\addlinespace
Cloud distribution & \$2.16B sales, 47\% of revenue;
roughly \$351M fees, FY2025 \citep{reuters2026anthropicdistribution}. &
Gross revenue recognition; fees enter operating expenses. Deduct them
once when calculating cash generation.\\
\addlinespace
Customer dependence & Two customers each supplied 12\% of FY2025
revenue; many large clients lacked long-term contracts
\citep{reuters2026anthropicdistribution}. &
Concentration and spending flexibility qualify the durability of demand.\\
\addlinespace
Adjusted operating income & Expected positive for a second
consecutive quarter, reported September 13, 2026
\citep{reuters2026anthropicadjustedprofit}. &
Evidence of reported improvement; exact adjustments and reconciliation
to free cash flow unavailable in the accessible report.\\
\addlinespace
Gross margin & Reportedly above 80\% before distribution sharing
and model training, September 2026
\citep{reuters2026anthropicadjustedprofit}. &
Partial cost coverage; neither a fully loaded cash margin nor a definition
of the separate adjusted operating-income measure.\\
\addlinespace
Infrastructure obligations & At least \$518B over about a decade;
about 80\% noncancelable or payable regardless of usage, reported
September 29, 2026 \citep{reuters2026anthropiccommitments}. &
Capacity access and less flexible purchases; neither annual expenditure
nor a net cash-deficit forecast.\\
\end{longtable}
\endgroup

\noindent\textit{Notes.}
Nominal US dollars. Annual payments, incremental revenue, taxes, cash investment,
working-capital changes, and a reconciliation of reported earnings would be needed to construct a comparable cash-flow path.

The reported compute and infrastructure spending was about 159\% of
2025 revenue, calculated as $7.33/4.6$ from rounded amounts \citep{reuters2026anthropicfinancials}.
This illustrates the scale of investment and costs relative to historical sales, but the aggregate does not separate model training from serving
customers and cannot establish the margin on additional usage. 
The later reported gross margin above 80\% excluded distribution sharing and model training \citep{reuters2026anthropicadjustedprofit}. It therefore gives a partial account of costs, rather than evidence for the fully loaded mature cash margin.

The numerical cases remain conditional tests of growth and cash retention.
A comparable Anthropic loss stress would require a dated net cash-flow
forecast or a reconciliation from earnings to cash investment, taxes, and working capital. 
Neither historical operating loss nor the multi-year purchase total supplies that path.
Capacity purchases and distribution fees are covered by the assumed cash margin. 

\Needspace{8\baselineskip}
\subsection{Subscription and API allocation}\label{app:allocation}

We examined how subscriptions, business seats, and separately billed API use
could supply the required year-ten revenue.
For consumer subscriptions $C_T$, annual receipts per subscription
$a_T$, business seats $B_T$, annual receipts per seat
$b_T$, and other revenue $O_T$, the remaining API condition is
$A_T=\max\{0,R_T-C_Ta_T-B_Tb_T-O_T\}$. The worked example sets
$O_T=0$. Advertising, non-access implementation, and other products
would enter $O_T$; unconsolidated investment returns do not enter this revenue identity. The exercise allocates an endpoint rather than forecasting each channel's growth or valuing its cash flows separately.
Subscription-bundled inference is excluded from separately billed API receipts. 
Token billing distinguishes fresh inputs, cache reads/writes, and output without overlapping quantities \citep{openrouter2026caching}.

Consumer plan shares $w_G,w_+,w_P$ sum to one. January US price benchmarks give $a=12(8w_G+20w_++200w_P)$ \citep{openai2026go}. 
All-Go gives \$96 annually, two-thirds Go/one-third Plus \$144, a 50\%
Go/49\% Plus/1\% Pro mix \$189.60, and all-Plus \$240. 
These are assumed mixes and list prices, not global realized receipts. 
Localization, discounts, taxes, and channel deductions matter; Pro subscriptions can raise the average above a Go versus Plus comparison. 
At 200 million subscriptions and 100 million seats paying \$240 annually, the two-thirds Go/one-third Plus mix gives \$52.8 billion. 
All-Go/all-Plus give \$43.2 billion/\$72 billion, leaving \$1.249-1.278 trillion in reference API receipts or \$275-304 billion in the lower setting.

The assumed counts are four times the 50-million lower bound for consumer subscribers and roughly eleven times the 9 million paying business users disclosed in February \citep{openai2026scale}. 
The \$240 benchmark reflects the \$20 monthly seat price under annual billing announced in April, reduced from \$25 \citep{openai2026businesspricing}.
The 216 allocation combinations vary consumer counts, seats, business receipts, and plan mix. 
Consumer and employer subscriptions may cover the same person only if both generate separate payments.

The gateway example uses rounded Vercel disclosures. 
Total volume rose 59\% and OpenAI's share rose from 10.3\% to 12.8\%, giving $1.59(0.128/0.103)=1.976$. Multiplying by the reported average-price factor 0.58 gives $1.146$ times list-price spending \citep{vercel2026july,vercel2026august}. 
This values gateway traffic, including customer-supplied keys, at published prices. 
It is neither measured OpenAI revenue nor representative worldwide demand, and it does not
identify a ten-year growth rate.

\subsection{Spending and capture comparisons}\label{app:spending}

To illustrate the spending shares needed to support the announced valuations, we consider hypothetical worldwide customer spending on AI software and services of \$2 trillion, \$5 trillion, or \$10 trillion annually in 2036.\footnote{For comparison, Gartner forecasts \$1.038 trillion of AI software and services spending in 2026 \citep{gartner2026ai}.
Its market categories differ from ours, and its figures may include
transactions between businesses. They therefore provide a rough
comparison of scale rather than a directly comparable measure
of final-customer spending.} 
Table~\ref{tbl:capture-frontier} shows the percentage of each total
that would need to reach OpenAI and Anthropic to meet their combined
revenue requirement under the different financial assumptions. 
For example, the reference requirement of \$2.791 trillion would absorb 55.8\% of a \$5 trillion total. 
A share above 100\% means that the assumed spending total is too small
to supply the required revenue. 

The comparison counts payments by final customers, including payments
for completed work as well as software access. 
We exclude sales between model providers to avoid counting the same spending more than once. 
We also assume that all required revenue comes from this spending pool,
with no additional receipts from sources such as advertising.

Figure~\ref{fig:feasibility} compares all three spending pools across 
mature cash margins of 10-40\% and discount rates of 10\% and 15\%.
In each case, the cash margin starts at -10\% and rises linearly to
the selected mature margin over ten years. 

\Needspace{20\baselineskip}

\begin{table}[htbp]
\centering
\caption{Required revenue and shares of hypothetical customer spending}\label{tbl:capture-frontier}
\begin{tabular}{@{}
  >{\raggedright\arraybackslash}p{\dimexpr 0.2000\linewidth - 1.6000\tabcolsep\relax}
  >{\raggedleft\arraybackslash}p{\dimexpr 0.2000\linewidth - 1.6000\tabcolsep\relax}
  >{\raggedleft\arraybackslash}p{\dimexpr 0.2000\linewidth - 1.6000\tabcolsep\relax}
  >{\raggedleft\arraybackslash}p{\dimexpr 0.2000\linewidth - 1.6000\tabcolsep\relax}
  >{\raggedleft\arraybackslash}p{\dimexpr 0.2000\linewidth - 1.6000\tabcolsep\relax}@{}}
\toprule\noalign{}
\begin{minipage}[b]{\linewidth}\raggedright
Discount rate; mature cash margin
\end{minipage} & \begin{minipage}[b]{\linewidth}\raggedleft
Joint required year-ten revenue
\end{minipage} & \begin{minipage}[b]{\linewidth}\raggedleft
\$2T pool: required joint share
\end{minipage} & \begin{minipage}[b]{\linewidth}\raggedleft
\$5T pool: required joint share
\end{minipage} & \begin{minipage}[b]{\linewidth}\raggedleft
\$10T pool: required joint share
\end{minipage} \\
\midrule\noalign{}

15\%; 15\% & \$4.846T & 242.3\% & 96.9\% & 48.5\% \\
15\%; 25\% (reference) & \$2.791T & 139.5\% & 55.8\% & 27.9\% \\
15\%; 40\% & \$1.677T & 83.8\% & 33.5\% & 16.8\% \\
10\%; 40\% (optimistic combination) & \$0.733T & 36.6\% & 14.7\% &
7.3\% \\
\bottomrule
\end{tabular}
\noindent\begin{minipage}{\linewidth}
\footnotesize
\textit{Notes.} Targets are OpenAI's \$852 billion and Anthropic's
\$965 billion reported valuations. Revenue is in trillions of nominal
US dollars; shares divide joint annual revenue by the stated worldwide
2036 final-customer spending pool. There are
no interprovider sales or outside-pool receipts. The first three rows
change only the mature margin; the fourth also changes the discount
rate. All use ten-year geometric revenue paths, linear margins from
-10\%, 2\% continuing growth, and $\lambda=q=1$.
\end{minipage}
\end{table}

\end{document}